\documentclass[a4paper,oneside]{amsart}

\usepackage{color}
\usepackage[metapost,centeredcaptions]{mfpic}
\RequirePackage{amsmath}
\RequirePackage{bm}
\RequirePackage{amssymb}
\RequirePackage{upref}
\RequirePackage{amsthm}
\RequirePackage{enumerate}
\RequirePackage{pb-diagram}
\RequirePackage{amsfonts}
\RequirePackage[mathscr]{eucal}
\RequirePackage{verbatim}
\RequirePackage{xr}
\RequirePackage{graphicx}
\usepackage{calc}
\usepackage{xspace}

\newcommand{\cf}{cf.\@\xspace}
\newcommand{\resp}{resp.\@\xspace}

\newcommand{\al}{\alpha}
\newcommand{\bet}{\beta}
\newcommand{\ga}{\gamma}
\newcommand{\de}{\delta }
\newcommand{\e}{\epsilon}

\newcommand{\f}{\varphi}

\newcommand{\ka}{\kappa}
\newcommand{\lam}{\lambda}

\newcommand{\om}{\omega}

\newcommand{\s}{\sigma}
\newcommand{\x}{\xi}

\newcommand{\C}{\varGamma}
\newcommand{\D}{\varDelta}
\newcommand{\F}{\varPhi}
\newcommand{\Lam}{\varLambda}
\newcommand{\Om}{\varOmega}

\newcommand{\so}{{\mc S_0}}

\newcommand{\const}{\tup{const}}

\newcommand{\msp[1]}[1]{\mspace{#1mu}}

\newcommand{\R}[1][n+1]{{\protect\mathbb R}^{#1}}

\newcommand{\N}{{\protect\mathbb N}}

\newcommand{\Z}{{\protect\mathbb Z}}
\newcommand{\eR}{\stackrel{\lower1ex \hbox{\rule{6.5pt}{0.5pt}}}{\msp[3]\R[]}}
\newcommand{\eN}{\stackrel{\lower1ex \hbox{\rule{6.5pt}{0.5pt}}}{\msp[1]\N}}
\newcommand{\eO}{\stackrel{\lower1ex
\hbox{\rule{6pt}{0.5pt}}}{\msc O}}

\newcommand\ra{\rightarrow}

\newcommand\pa{\partial}

\newcommand{\un}{\infty}
\newcommand{\A}{\forall}

\newcommand{\set}[2]{\{\,#1\colon #2\,\}}
\newcommand{\uu}{\cup}

\newcommand{\uuu}{\bigcup}

\newcommand{\uud}{ \stackrel{\lower 1ex \hbox {.}}{\uu}}
\newcommand{\uuud}[1]{ \stackrel{\lower 1ex \hbox {.}}{\uuu_{#1}}}
\newcommand\su{\subset}

\newcommand{\sminus}[1][28]{\raise 0.#1ex\hbox{$\scriptstyle\setminus$}}

\newcommand\ti{\times }

\newcommand{\abs}[1]{\lvert#1\rvert}

\newcommand{\norm}[1]{\lVert#1\rVert}

\newcommand{\spd}[2]{\protect\langle #1,#2\protect\rangle}

\newcommand\cha[3]{{\bar\varGamma}_{#1#2}^#3}

\newcommand{\tit}{\textit}

\newcommand{\tup}{\textup}% text upright

\newcommand{\mc}{\protect\mathcal}
\newcommand{\msc}{\protect\mathscr}

\newcommand{\cq}[1]{\glqq{#1}\grqq\,}

\newcommand{\bt}{\begin{thm}}
\newcommand{\bl}{\begin{lem}}
\newcommand{\bc}{\begin{cor}}
\newcommand{\bd}{\begin{definition}}
\newcommand{\bpp}{\begin{prop}}
\newcommand{\br}{\begin{rem}}
\newcommand{\bn}{\begin{note}}
\newcommand{\be}{\begin{ex}}
\newcommand{\bes}{\begin{exs}}
\newcommand{\bb}{\begin{example}}
\newcommand{\bbs}{\begin{examples}}
\newcommand{\ba}{\begin{axiom}}

\newcommand{\et}{\end{thm}}
\newcommand{\el}{\end{lem}}
\newcommand{\ec}{\end{cor}}
\newcommand{\ed}{\end{definition}}
\newcommand{\epp}{\end{prop}}
\newcommand{\er}{\end{rem}}
\newcommand{\en}{\end{note}}
\newcommand{\ee}{\end{ex}}
\newcommand{\ees}{\end{exs}}
\newcommand{\eb}{\end{example}}
\newcommand{\ebs}{\end{examples}}
\newcommand{\ea}{\end{axiom}}

\newcommand{\bp}{\begin{proof}}
\newcommand{\ep}{\end{proof}}
\newcommand{\eps}{\renewcommand{\qed}{}\end{proof}}

\newcommand{\bal}{\begin{align}}

\newcommand{\bi}[1][1.]{\begin{enumerate}[\upshape #1]}
\newcommand{\bia}[1][(1)]{\begin{enumerate}[\upshape #1]}
\newcommand{\bin}[1][1]{\begin{enumerate}[\upshape\bfseries #1]}
\newcommand{\bir}[1][(i)]{\begin{enumerate}[\upshape #1]}
\newcommand{\bic}[1][(i)]{\begin{enumerate}[\upshape\hspace{2\cma}#1]}
\newcommand{\bis}[2][1.]{\begin{enumerate}[\upshape\hspace{#2\parindent}#1]}
\newcommand{\ei}{\end{enumerate}}

\newcommand\ndots{\raise 0.47ex \hbox {,}\hskip0.06em\cdots %
     \raise 0.47ex \hbox {,}\hskip0.06em} 

\newcommand{\q}{\quad}
\newcommand{\qq}{\qquad}

\newcommand{\hp}{\hphantom}

\newcommand\nd{\noindent}

\newskip\Csmallskipamount                                                
\Csmallskipamount=\smallskipamount
\newskip\Cmedskipamount
\Cmedskipamount=\medskipamount
\newskip\Cbigskipamount
\Cbigskipamount=\bigskipamount

\newcommand\cvs{\vspace\Csmallskipamount}   
\newcommand\cvm{\vspace\Cmedskipamount}
\newcommand\cvb{\vspace\Cbigskipamount}

\newskip\csa
\csa=\smallskipamount

\newskip\cma
\cma=\medskipamount

\newskip\cba
\cba=\bigskipamount

\newdimen\spt
\newcommand\citem{\cvs\advance\itemno by
1{(\romannumeral\the\itemno})\hskip3pt}
\newcommand{\bitem}{\cvm\nd\advance\itemno by
1{\bf\the\itemno}\hspace{\cma}}

\newcount\itemno
\newcommand{\las}[1]{\label{S:#1}}

\newcommand{\lae}[1]{\label{E:#1}}
\newcommand{\lat}[1]{\label{T:#1}}
\newcommand{\lal}[1]{\label{L:#1}}

\newcommand{\lar}[1]{\label{R:#1}}

\newcommand{\rt}[1]{Theorem~\ref{T:#1}}
\newcommand{\rl}[1]{Lemma~\ref{L:#1}}

\newcommand{\rr}[1]{Remark~\ref{R:#1}}

\newcommand{\re}[1]{\eqref{E:#1}}

\newskip\thmskip
\thmskip=\parindent

\newskip\hsk
\newenvironment{hinw}{\labelsep=0pt\begin{list}{}{\labelsep=0pt\itemindent=0pt\labelwidth=0pt\leftmargin=\parindent\rightmargin=0pt\partopsep=\cba}%
\item\it\nopagebreak\nopagebreak}%
{\end{list}}

\newcommand\bh{\begin{hinw}}
\newcommand{\eh}{\end{hinw}}

\newtheoremstyle{normal}% name
  {\cba}%      Space above, empty = `usual value'
  {\cba}%      Space below
  {}% Body font
  {\thmskip}%Indent amount (empty = no indent, \parindent = para indent)
  {\bfseries}% Thm head font
  {.}%        Punctuation after thm head
  {\hsk}%     Space after thm head: " " = normal interword space;
  {}% Thm head spec

\newtheoremstyle{abschnitt}% name
  {\cba}%      Space above, empty = `usual value'
  {\cba}%      Space below
  {}% Body font
  {\thmskip}% Indent amount (empty = no indent, \parindent = para indent)
  {\bfseries}% Thm head font
  {.}%        Punctuation after thm head
  {\hsk}%     Space after thm head: " " = normal interword space;
  {}% Thm head spec

\newtheoremstyle{italic}% name
  {\cba}%      Space above, empty = `usual value'
  {\cba}%      Space below
  {\itshape}% Body font
  {\thmskip}%  Indent amount (empty = no indent, \parindent = para indent)
  {\bfseries}% Thm head font
  {.}%        Punctuation after thm head
  {\hsk}%     Space after thm head: " " = normal interword space;
  {}% Thm head spec

\newtheoremstyle{aufgaben}% name
  {\cba}%      Space above, empty = `usual value'
  {\cba}%      Space below
  {}% Body font
  {}%         Indent amount (empty = no indent, \parindent = para indent)
  {\normalsize\bfseries}% Thm head font
  {.}%        Punctuation after thm head
  {\hsk}%     Space after thm head: " " = normal interword space;
  {}% Thm head spec

\newtheoremstyle{break}% name
  {\cba}%      Space above, empty = `usual value'
  {\cba}%      Space below
  {\itshape}% Body font
  {}%         Indent amount (empty = no indent, \parindent = para indent)
  {\bfseries}% Thm head font
  {.}%        Punctuation after thm head
  {\newline}% Space after thm head: \newline = linebreak
  {}%         Thm head spec

\swapnumbers
\theoremstyle{italic}
\newtheorem{thm}[subsection]{Theorem}
\newtheorem{lem}[subsection]{Lemma}
\newtheorem{prop}[subsection]{Proposition}
\newtheorem{cor}[subsection]{Corollary}

\theoremstyle{normal}
\newtheorem{rem}[subsection]{Remark}
\newtheorem{definition}[subsection]{Definition}
\newtheorem{example}[subsection]{Example}
\newtheorem{examples}[subsection]{Examples}
\newtheorem{ex}[subsection]{Exercise}
\newtheorem{note}[subsection]{}
\newtheorem{axiom}[subsection]{Axiom}

\theoremstyle{aufgaben}
\newtheorem{exs}[subsection]{Exercises}

\swapnumbers

\numberwithin{equation}{section}
\numberwithin{figure}{section}

\newenvironment{textequation}[1][0.8]
{\begin{equation}
\begin{aligned}
\begin{minipage}{#1\linewidth}}
{\end{minipage}
\end{aligned}
\end{equation}
\ignorespacesafterend}

\newcommand{\btext}{\begin{textequation}}
\newcommand{\etext}{\end{textequation}}

\newcommand{\si}[1]{_{{\vphantom{}}_{#1}}}

\usepackage[german,english]{babel}
\usepackage{graphicx}
\RequirePackage{amsmath}
\RequirePackage{bm}
\RequirePackage{amssymb}
\RequirePackage{upref}
\RequirePackage{amsthm}
\RequirePackage{enumerate}
\RequirePackage{pb-diagram}
\RequirePackage{amsfonts}
\RequirePackage[mathscr]{eucal}
\RequirePackage{verbatim}
\RequirePackage{xr}
\RequirePackage{graphicx}
\RequirePackage[pdftex]{floatflt}
\usepackage{calc}
\usepackage{xspace}
\newcommand{\fried}{\ensuremath{(\tilde h+\tfrac{\ka^2}{n^2}(V+\s+\mc
C)^2r^2)}}

\newcommand{\friedc}{\ensuremath{\hat h+\tfrac{\ka^2}{n^2}(\rho r^{2n}+\s
r^{2n}+\mc C r^{2n})^2}}

\newcommand{\frcc}{\ensuremath{\hat
h+\tfrac{\ka^2}{n^2}(\rho r^{2n}+\mc{\hat C})^2}}

\newcommand{\dis}{\ensuremath{\sqrt{1-2\norm{\dot\f}^2
\tfrac{\ka^2}{n^2}(V+\s+\mc C)r^2 -\tfrac{\ka^2}{n^2}\norm{\dot\f}^4\tilde
hr^2}}}

\newcommand{\kn}{\ensuremath{\tfrac{\ka^2}{n^2}}}

\RequirePackage{upref}
\RequirePackage{amsthm}
\RequirePackage{enumerate}%\begin{enumerate}[(i)]
\usepackage[mathscr]{eucal}
\usepackage{xr-hyper}
\usepackage{hyperref}

\usepackage{calc}

\newlength{\oddsidemarginlength}
\newlength{\topmarginlength}

\newcounter{numberoflines}
\newcounter{tempcc}
\oddsidemargin=\oddsidemarginlength
\evensidemargin=\oddsidemargin
\topmargin=\topmarginlength
\opengraphsfile{brane2}

\headshape{1}{1}{true}

\renewcommand\raggedright{\rightskip 0pt plus2em \spaceskip.3333em \xspaceskip.5em\relax}

\begin{document}

\flushbottom

%\larger[1]
%\frontmatter

\title[Branes and moduli spaces]{Branes, moduli spaces and smooth transition
from big crunch to big bang}

% author one information
\author{Claus Gerhardt}
\address{Ruprecht-Karls-Universit\"at, Institut f\"ur Angewandte Mathematik,
Im Neuenheimer Feld 294, 69120 Heidelberg, Germany}
%\curraddr{}
\email{gerhardt@math.uni-heidelberg.de}
\urladdr{http://www.math.uni-heidelberg.de/studinfo/gerhardt/}
\thanks{This work has been supported by the Deutsche Forschungsgemeinschaft.}

% author two information
%\author{}
%\address{}
%\curraddr{}
%\email{}
%\thanks{}
%
\subjclass[2000]{35J60, 53C21, 53C44, 53C50, 58J05}
\keywords{Lorentzian manifold, branes, moduli spaces, general relativity, transition from big crunch to big bang}
\date{\today}
%
% at present the "communicated by" line appears only in ERA and PROC
%\commby{}

%\dedicatory{}

\begin{abstract}
We consider branes $N$ in a Schwarzschild-$\text{AdS}_{(n+2)}$ bulk, where the
stress energy tensor is dominated by  the energy density of  a scalar fields map
$\f:N\ra
\mc S$ with potential $V$, where $\mc S$ is a semi-Riemannian moduli space. By
transforming the field equation appropriately, we get an equivalent field equation
that is smooth across the singularity $r=0$, and which has smooth and uniquely
determined solutions which exist across the singularity in an interval $(-\e,\e)$.
Restricting a solution to $(-\e,0)$ \resp $(0,\e)$, and assuming $n$ odd, we
obtain branes $N$ \resp $\hat N$ which together form a smooth hypersurface. Thus
a smooth  transition from big crunch to big bang is possible both geometrically as well
as physically.
\end{abstract}

\maketitle

\tableofcontents

\setcounter{section}{0}
\section{Introduction}

Finding a smooth transition from big crunch to big bang is an interesting and
challenging problem in general relativity. For abstract spacetimes, i.e., for spacetimes
that are not embedded in a bulk space, it is even a non-trivial question how to define
a smooth transition. Using the inverse mean curvature flow in combination with a
reflection we proved in \cite{cg:arw,cg:rw} that a smooth transition is possible for a
certain class of spacetimes which we called ARW spaces.

For branes the phrasing of the problem is simpler and also the proof that there are
solutions to the problem is much easier---at least, if the stress-energy tensor on the
brane is made up of perfect fluids satisfying an equation of state with constant
$\om$, and of course, under the general assumption of radial symmetry,
\cf
\cite{cg:brane}.

If scalar fields with non-trivial potentials are considered in the stress-energy tensor,
then the problem is much harder to solve, since the $\om$ in the equation of state is
not constant, and hence, the conservation law cannot be easily integrated.

\cvm
We look at branes $N$ in a Schwarzschild-$\tup{AdS}_{(n+2)}$ bulk $\mc
N=\R[]_-\times\R[]\times\so$, where $\so$ is an $n$-dimensional space form,
assuming that the branes are located in the black hole region.

The relation between the geometry of the brane and physics is governed by the Israel
junction condition.

Let $\Om\su\mc N$ be an open set contained in the black hole region, which is 
bounded by (part of) the horizon and two branes (timelike hypersurfaces) $N_i$,
$i=1,2$, where $N_2$ may be empty.

The field equations are derived from an action principle, where the action integral $I$
is given by
\begin{equation}\lae{0.1}
I=\int_\Om (\tilde R - 2\Lam)+\sum_{i=1}^2\int_{N_i}(2 H_i+4\ka(-\s_i+
\sideset{_{i}}{}{\operatorname {\it L}_{\msp tot}})).
\end{equation}
$\tilde R$ is the scalar curvature of $\mc N$, $\Lam$ a cosmology constant, $H_i$
the
\resp mean curvature of the
$N_i$, with respect to their \tit{inner normal}, i.e., with respect to the normal that is
pointing towards $\Om$, 
$\s_i$ their tensions, and
$\sideset{_{i}}{}{\operatorname {\it L}_{\msp tot}}$ the
\resp Lagrange functions corresponding to $N_i$; the Lagrange functions may be
different, but are assumed to have the same structure, so that it suffices to described
only the case
$i=1$ in detail, where we omit the index $1$ dealing instead with a generic brane $N$.
$L_{tot}$ is then defined by
\begin{equation}
L_{tot}=\sum_{k=0}^{k_0}L_k+(-\tfrac12\norm{D\f}^2-V(\f)),
\end{equation}
where $L_k$ are Lagrange functions corresponding to perfect fluids satisfying an
equation of state with constant $\om_k$
\begin{equation}\lae{0.3}
p_k=\tfrac{\om_k}{n}\rho_k
\end{equation}
and $\f=(\f^A)$ is a map
\begin{equation}
\f:N\ra \mc S,
\end{equation}
where the \tit{moduli space} $\mc S$ is  an arbitrary semi-Riemannian
manifold of finite dimension with metric $(g\si{AB})$ and $V$ a smooth
potential.\footnote{Note that $\mc S$ and $V$ can in general be different for
different branes.} The Lagrange function
\begin{equation}\lae{0.5}
L=-\tfrac12\norm{D\f}^2-V(\f)=-\tfrac12g\si{AB}\f^A_\al\f^B_\bet\bar
g^{\al\bet}-V(\f)
\end{equation}
can be looked at as the energy density of the map $\f$.

\cvm
The \tit{power law} Lagrangians $L_k$ are only included for good measure, because
they have already been treated in \cite{cg:brane} and pose only a slight additional
challenge as long as the $\om_k\in\Z$ are bounded by
\begin{equation}
\om_k\le n.
\end{equation}

The only interesting Lagrangian is the one given in \re{0.5}.

\cvm
In our previous paper, where we treated power law Lagrangians, we first solved the
embedding problem in $\mc N$, then reflected the bulk space by switching the light
cone and changing the radial coordinate from $r$ to $-r$ to obtain a new bulk $\mc
{\hat N}$ which had a white hole singularity in $r=0$. The spaces $\mc N$ and $\mc
{\hat N}$ could be pasted together to yield $\R[2]\times \so$ topologically.

Then we used a reflection method to create a reflected brane in $\hat N\su \mc
{\hat N}$ such that $N\uu\hat N$ would form a smooth hypersurface in
$\R[2]\times \so$.

\cvm
In the present case, where the stress-energy tensor corresponding to the scalar
fields map will dominate the evolution of the brane near the singularity, a much
better, if not optimal, result can be proved.

\cvm
By choosing the radial coordinate as embedding parameter we are able to transform
the field equation of the brane near the singularity to obtain a new equation which is
smooth across the singularity and has smooth solutions which are uniquely
determined by given initial values at the singularity. 
 
This new field equation has
 to be satisfied near the singularity by any solution of the embedding problem provided
its rescaled energy density $\rho r^{2n}$ has a positive limit when the singularity is
approached and the limit of the rescaled potential $Vr^{2n}$ vanishes.

\cvm
If we assume that $n$ and the $\om_k$ are odd, then we can show  that the
abstract solution of the new field equation is actually a physical solution on both sides
of the singularity, and it exists in a small interval $\{-\e<r<\e\}$.

\cvm
The action integrals $I$ in $\mc N$ \resp $\mc {\hat N}$ will be identical with the
exception that, if we decompose a (globally) smooth brane in its parts $N\su\mc N$
and $\hat N\su \mc{\hat N}$ and if $\Om$ \resp $\hat \Om$ are the corresponding
domains, then $N$ \resp $\hat N$ have opposite positions with respect to their
domains, e.g., if $\Om$ lies below $N$, in which case $N$ is called an \tit{upper}
brane, then $\hat N$ is a \tit{lower} brane, i.e., $\hat \Om$ lies above $\hat N$.
Notice that the branes are graphs so that the $t$-axis provides a direction.

\cvm
This behaviour is also responsible for the fact that a configuration with two boundary
branes has a smooth \cq{mirror} image in $\mc {\hat N}$, if and only if the branes
collide at the singularity and switch positions in $\mc {\hat N}$.

\cvm
It is remarkable that a smooth transition from big crunch to big bang is possible
without additional symmetry assumptions on the potential or on the moduli space.

However, if $\mc S$ is locally symmetric and if there exists a point $p\in \mc S$
such that $V$ is invariant in a neighbourhood of $p$ with respect to the geodesic
symmetry at $p$, then, by choosing $p$ as one initial value at the singularity, it can
be shown that the solutions in $\mc N$ \resp $\mc {\hat N}$ correspond to each
other via reflection near the singularity, after having introduced Riemannian normal coordinates at $p$.

\cvm
Moreover, if $\mc S$ is Riemannian and  complete and the energy density $\rho$
bounded from below, then we can prove that the branes extend to the \resp horizons,
i.e., they are defined in $(-r_0,r_0)$, where $r_0$ is the black (white) hole radius.

At the horizon our embedding experiences a coordinate singularity. Thus, an
extension past the horizon seems possible but we didn't pursue this point.

\cvm
In order to describe our results accurately, let us introduce some definitions and
notations. 

The metric in the bulk space $\mc N$ is given by 
\begin{equation}\lae{0.7}
d\tilde s^2=-\tilde h^{-1} dr^2 +\tilde hdt^2 + r^2\s_{ij}dx^idx^j,
\end{equation}
where $(\s_{ij})$ is the metric of an $n$- dimensional space form $\so$, the radial
coordinate
$r$ is assumed to be negative, $r<0$, and $\tilde h(r)$ is defined by
\begin{equation}
\tilde h=m(-r)^{-(n-1)}+\tfrac2{n(n+1)}\Lam r^2 -\tilde\ka,
\end{equation}
where $m>0$ and $\Lam\le 0$ are constants, and $\tilde\ka=-1,0,1$ is the
curvature of $\so$. We note that we assume that there is a black hole region, i.e., if
$\Lam=0$, then we have to suppose $\tilde\ka =1$.

\cvm
We consider branes $N$ contained in the black hole region $\{-r_0<r<0\}$ given by
an
embedding
\begin{equation}\lae{0.9}
y(\tau, x^i)=(r(\tau),t(\tau),x^i),\q -a<\tau<0,
\end{equation}
with a big crunch singularity in $\tau=0$ such that $r(0)=0$.
$N$ will be a globally hyperbolic spacetime $N=I\ti \so$ with metric $(\bar
g_{\al\bet})$ that can be expressed in the so-called conformal time gauge as
\begin{equation}\lae{0.10}
d\bar s^2=e^{2f}(-(dx^0)^2+\s_{ij}dx^idx^j)
\end{equation}
such that
\begin{equation}
f=f(x^0)=\log(-r(x^0)).
\end{equation}

\cvm
Another gauge is even more important for our purposes, namely, by choosing
$x^0=r$ as time function. The embedding  then has the form
\begin{equation}\lae{0.12}
y(r, x^i)=(r,t(r),x^i), \q r<0,
\end{equation}
and the metric is given by
\begin{equation}\lae{0.13}
d\bar s^2= -(\tilde h^{-1}-\tilde h\abs{t'}^2) dr^2 +r^2 \s_{ij}dx^idx^j.
\end{equation}

\cvm
The relation between geometry and physics is governed by the Israel junction
conditions
\begin{equation}\lae{0.14}
h_{\al\bet}-H\bar g_{\al\bet}=\ka (T_{\al\bet} -\s\bar g_{\al\bet}),
\end{equation}
where $h_{\al\bet}$ is the second fundamental form of $N$,
$H=\bar g^{\al\bet}h_{\al\bet}$ the mean curvature,
$\ka\ne 0$ a constant,
 $T_{\al\bet}$ the stress energy tensor
 and $\s$ the tension of the brane.

\cvm
Let $\om_k$, $0\le k\le k_0$, be the constants in \re{0.3}, where $\om_0$ is
assumed to satisfy
\begin{equation}
\om_0=n,
\end{equation}
and let $0\le \bar \rho_k$ be the corresponding integration constants. Define
\begin{equation}
\mc C=\sum_{k=0}^{k_0}\bar\rho_k (-r)^{-(n+\om_k)},\q \mc {\hat C}=(\s+\mc C)
r^{2n},\q \hat h=\tilde h r^{2n},
\end{equation}
and
\begin{equation}\lae{0.17}
D^2=1-2\kn\norm{\dot\f}^2(V+\s+\mc C)r^{2n}-\kn\tilde h\norm{\dot\f}^4.
\end{equation}

\cvm
The main results of this paper can now be summarized in the following three
theorems.

\bt\lat{0.1}
Let $\mc S$ be semi-Riemannian with metric $(g\si{AB})$ and assume that the
metric has at least one positive eigenvalue. The potential $V$ should be of class
$C^\un(\mc S)$. 

\cvm
\tup{(i)} Consider the action integral \re{0.1} for a single brane and look at
embeddings of the form \re{0.12} parametrized by $r$ with a corresponding scalar
fields map $\f(r)$. The field equations for $\f$ resulting from the Israel junction
conditions and the zero divergence of the stress-energy tensor associated with $\f$,
can be combined to a new field equation of the form
\begin{equation}\lae{0.18}
\frac D{dr}\chi^A=F^A(r,\f,\chi),
\end{equation}
where
\begin{equation}
\chi^A=\dot\f r^{-(n-1)},
\end{equation}
provided the term $D^2$ in \re{0.17} is strictly positive.

Treating the arguments of $F^A$ as independent variables $(r,\f,\chi)$, there holds
$F^A\in C^\un((-\e,\e)\times \mc S \times T^{1,0}(\mc S))$ as long as
$\norm{\chi}^2>0$ and $\e$ is small, more precisely,
\begin{equation}\lae{0.20}
\begin{aligned}
F^A&=-\frac{\hat h'
+2\kn (\rho r^{2n}+\mc {\hat C})(2nV r^{2n-1}+\mc{\hat
C}')}{\frcc}\chi^A\\[\cma] 
&\hp{=}-\frac{\pa V}{\pa \f\si A}
\,\frac{r^{3n-1}}{\frcc},
\end{aligned}
\end{equation}
where the term $\rho r^{2n}$ can also be expressed by $(r,\f,\chi)$.

These field equations have to be satisfied near the  singularity by any solution of
the embedding problem that also solves the wave equation \re{2.10}, provided the
corresponding quantities
$\rho r^{2n}$ and
$Vr^{2n}$ obey
\begin{equation}
\lim_{r\ra 0}\rho r^{2n}>\bar\rho_0
\end{equation}
and
\begin{equation}
\lim_{r\ra 0}Vr^{2n}=0.
\end{equation}

\cvm
\tup{(ii)} The field equations \re{0.18} can be solved abstractly in a full
neighbourhood of the singularity $r=0$ with given initial values $\f(0)$, $\chi(0)$,
provided $\norm{\chi(0)}^2>0$ and
\begin{equation}
2\kn \norm{\chi(0)}^2\bar\rho_0<1,
\end{equation} 
for details we refer to \rt{2.4}. The solutions $(\f,\chi)$ are smooth in $(-\e,\e)$ and
are related by
\begin{equation}
\dot\f=\chi r^{n-1}.
\end{equation}
If we restrict $\f$ to $(-\e,0)$, then it is a solution of the original embedding
problem in $\mc N$. The function $t(r)$ is smooth in $(-\e,0]$. This result holds for
arbitrary $n$.

\cvm
\tup{(iii)} If in addition $n$ and the $\om_k$'s are odd, then the globally defined
abstract solution in \tup{(i)}  also defines an embedded brane $\hat N\su\mc{\hat
N}$. If $N$ is an upper brane, then $\hat N$ is a lower brane and vice versa.
The joint branes form a smooth hypersurface in $\R[2]\times \so$ which is a graph
$t=t(r)$, $t\in C^\un((-\e,\e))$. 

\cvm
\tup{(iv)} If $\mc S$ is Riemannian and complete, and $\rho$ bounded from below,
then the branes $N$ \resp $\hat N$ extend to the horizons. They probably extend
past the horizons, but the embeddings experience a coordinate singularity at the
horizons.
\et

\bt
If in addition to the assumptions in part \tup{(iii)} of the preceding theorem, $\mc S$
is locally symmetric, and there exists $p\in\mc S$ such that the potential $V$ is
invariant in a neighbourhood of $p$ with respect to the geodesic symmetry centered
at $p$, then, if $p$ is chosen as initial value in \tup{(iii)}, the solution $\f$, after the
introduction of Riemannian normal coordinates around $p$, is  odd in a neighbourhood
of $r=0$, and the brane $\hat N$ is the odd reflection of the brane $N$ near the singularity.
\et

\bt
Consider the action integral \re{0.3} with two boundary branes. If the assumptions of
part \tup{(iii)} of \rt{0.1} are satisfied, then the  configuration $(\Om,N_1,N_2)$
has a smooth correspondence $(\hat\Om,\hat N_1,\hat N_2)$ in $\mc{\hat N}$, only if
 the branes $N_1$ and $N_2$ collide at the singularity. They then switch
positions in $\mc{\hat N}$. 
If the assumptions of part \tup{(iii)} of \rt{0.1} are valid, then this condition is also
sufficient. Moreover, the requirement of colliding boundary branes at $r=0$ can be
easily met.
\et

\section{Domain walls}

The configuration in the previous section, with a domain $\Om\su\mc N$ bounded by
one or two branes, can be generalized by assuming that a brane $N$ separates two
subdomains $\Om_k\su\mc N_k$, $k=1,2$, where $\mc N_k$ are different bulk
spaces, and where it is assumed that the induced metric of the brane from either side
is the same. 

\cvm
In many papers it has been assumed that this configuration is possible for bulk spaces
and embeddings where the ambient metrics have the form as in \re{0.7} and the
embedding of the brane is given by \re{0.9}, \cf, e.g., \cite{kraus:brane}. We
shall show that this is not possible. The only allowed difference between the bulk
spaces in the described setting would be one caused by $\Z_2$ symmetry.

\bpp
Let $N$ be a brane separating two subdomains of 
Schwarz\-schild-$\text{AdS}_{(n+2)}$ bulk spaces
$\mc N_k$, $k=1,2$, with metrics
 given by 
\begin{equation}\lae{1.1}
d\tilde s^2_k=-\tilde h^{-1}_k dr^2 +\tilde h_kdt^2 + r^2\s_{ij}dx^idx^j,
\end{equation}
where $(\s_{ij})$ is the metric of an $n$- dimensional space form $\so$, the radial
coordinate
$r$  can be either negative or positive, and is allowed to have different signs in the
bulk spaces, and
$\tilde h_k(r)$ is defined by
\begin{equation}
\tilde h_k=m_k\abs{r}^{-(n-1)}+\tfrac2{n(n+1)}\Lam_k r^2 -\tilde\ka,
\end{equation}
where $m_k\ge 0$ and $\Lam_k$ are constants, and $\tilde\ka=-1,0,1$ is the
curvature of $\so$.  The induced metric on the
brane from both sides should be same. If we then assume that the embedding of $M$
is given by
\begin{equation}\lae{1.3}
y(\tau)=(\pm r(\tau),\pm t(\tau),x^i),\q r<0,
\end{equation}
where the plus sign corresponds to the case that the $r$ coordinate in $\mc N_1$ or
in $\mc N_2$ is negative, then there must hold
\begin{equation}
\tilde h_1=\tilde h_2,
\end{equation}
i.e., the only possible difference between $\mc N_1$ and $\mc N_2$ is a $\Z_2$
symmetry.
\epp

\bp
We have written the metric in a form that suits our particular assumption that the
branes are located  in the black hole region. However, this is no restriction, since we
do not assume any sign condition on
$\tilde h_k$, i.e., $\tilde h_k$ could be negative and the bulk spaces could be AdS or 
dS spaces; in fact only the structure of the metric in \re{1.1} is relevant with arbitrary
$\tilde h_k$---satisfying $\tilde h_k(r)=\tilde h_k(-r)$, if we want to include $\Z_2$
symmetry---, without any assumptions on the curvature.

\cvm
The induced metric can be written as
\begin{equation}
d\bar s^2=r^2(-(dx^0)^2+\s_{ij}dx^idx^j).
\end{equation}

We also assume that $\dot r>0$. Notice that  $r<0$, so that $(x^\al)$ is a
future oriented coordinate system on the brane. 

Let us point out that this choice of $\tau$ implies the relation
\begin{equation}\lae{1.7}
r^2=\tilde h_k^{-1}\dot r^2 -\tilde h_k\abs{t'}^2,
\end{equation}
since
\begin{equation}
\bar g_{00}= \spd{\dot y}{\dot y},
\end{equation}
or equivalently,
\begin{equation}\lae{1.8}
\abs{\dot r}^2=\tilde h_kr^2+\tilde h_k^2\abs{t'}^2.
\end{equation}

\cvm
We assume furthermore that there is an open interval $I$ such that $y(\tau)$, for
$\tau\in I$, separates corresponding regions in $\mc N_k$, i.e., $y(\tau)$ is either
inside the event horizons---if they exist---or outside. Without loss of generality let us
suppose that we are inside the black hole regions, so that $\tilde h_k>0$, $k=1,2$,
otherwise we simply exchange the roles of $r$ and $t$, and write $h_k$ for $-\tilde
h_k^{-1}$ and $-h_k^{-1}$ for $\tilde h_k$.

\cvm
Subtracting then the equations \re{1.8} for $k=1$ \resp $k=2$, we get
\begin{equation}
0=(\tilde h_1-\tilde h_2)r^2+(\tilde h_1^2-\tilde h_2^2)\abs{t'}^2
\end{equation}
and, thus, if $\tilde h_1\ne \tilde h_2$,
\begin{equation}
0=r^2+(\tilde h_1+\tilde h_2)\abs{t'}^2,
\end{equation}
a contradiction. Hence, the equality $\tilde h_1=\tilde h_2$ must be valid.
\ep

If the bulk spaces $\mc N_k$ should really be different, aside from $\Z_2$
symmetry, which would be characterized by having different signs for $r$ in the bulk
spaces,  then the embeddings in
\re{1.3} must differ in more than a sign. Especially the equation \re{1.8} should look
differently for
$k=1$
\resp
$k=2$.

\cvm
Since $\Z_2$ symmetry would change the equation \re{0.14} only by a factor $2$,
our assumption that the branes are boundaries of a connected domain does not
constitute a lack of generality.

\section{The field equations}

Let us first summarize a few of our conventions. Covariant derivatives are always
\tit{full} tensors, i.e., for a map
\begin{equation}
\f: N\ra \mc S
\end{equation}
the Laplacian $\D\f^A$ is defined by  
\begin{equation}
\D\f^A=\bar g^{\al\bet}\f^A_{\al\bet},
\end{equation}
where $\f^A_{\al\bet}$
are second order covariant derivatives of $\f$ which are defined by
\begin{equation}\lae{2.3}
\f^A_{\al\bet}=\f^A_{,\al\bet} - \cha \al\bet\ga\f^A_\ga +\hat
\C^A_{BC}\f^B_\al\f^C_\bet,
\end{equation}
where a comma indicates as usual ordinary partial derivatives.

\cvm
Recall the Gau{\ss} formula
\begin{equation}\lae{2.4}
y_{\al\bet}=-\s h_{\al\bet}\msp \nu
\end{equation}
for a hypersurface $N\su \mc N$ given by an embedding $y=y(\x^\al)$, where
$\s=\spd\nu\nu$, which implicitly defines the second fundamental form. For a brane,
$\nu$ is spacelike, i.e., $\s=1$.

For a more detailed overview of our conventions we refer to \cite[Section
2]{cg:indiana}, where hypersurfaces in an albeit different setting are considered, but
this shouldn't cause any difficulties.

\cvm
Suppose now that the metric in the bulk and the embeddings of the boundary branes
satisfy a variational principle, namely, that the first variation of the functional $I$ in
\re{0.1} vanishes with respect to variations of the metric that have compact support
in $\Om\uu N_1\uu N_2$. Then, by first choosing variations with compact support in
$\Om$, we deduce that the bulk metric $(\tilde g_{ab})$ is an Einstein metric
\begin{equation}
G_{ab}+\Lam \tilde g_{ab}=0,
\end{equation}
and by looking at a general variation we obtain, after an integration by parts, as a free
boundary condition for each brane, the so-called Israel junction conditions
\begin{equation}\lae{2.6}
\begin{aligned}
h_{\al\bet}- H\bar g_{\al\bet}&=\ka \Big(L_{tot}\,\bar
g_{\al\bet}-2\frac{\pa L_{tot}}{\pa \bar g^{\al\bet}}-\s\bar
g_{\al\bet}\Big)\\[\cma]
&=\ka(\hat T_{\al\bet}-\s\bar g_{\al\bet}),
\end{aligned}
\end{equation}
where
\begin{equation}
\hat T_{\al\bet}=\sum_{k=0}^{k_0}\sideset{_{k}}{}{\operatorname{{\it
T_{\al\bet}}}}+ T_{\al\bet}
\end{equation}
is the total stress-energy tensor, $(\sideset{_{k}}{}{\operatorname{{\it
T_{\al\bet}}}})$, $0\le k\le k_0$, the stress-energy tensors of perfect fluids with
constant $\om_k$, $\om_k\in\Z$, and $(T_{\al\bet})$ the stress-energy tensor
associated with the scalar fields, i.e., associated with the map $\f$
\begin{equation}
T_{\al\bet}=(-\tfrac12 \norm{D\f}^2-V)\bar g_{\al\bet}
+g\si{AB}\f^A_\al\f^B_\bet.
\end{equation}

\cvm
Since the bulk metric, which we now assume to be of the form \re{0.7}, is Einstein,
the left-hand side of \re{2.6} is divergence free, in view of the Codazzi equations, i.e.,
$(\hat T_{\al\bet})$ is divergence free. Assuming in addition that each of the
$(\sideset{_{k}}{}{\operatorname{{\it T_{\al\bet}}}})$ has vanishing divergence,
then there must hold
\begin{equation}
0=T^\al_{\bet;\al}=(\D\f^A-\frac{\pa V}{\pa \f\si A}) \frac{\pa \f\si A}{\pa
x^\bet}.
\end{equation}

Though this does not necessarily imply that $\f$ satisfies the wave equation
\begin{equation}\lae{2.10}
\D\f^A-\frac{\pa V}{\pa \f\si A}=0,
\end{equation}
the reverse is certainly true.

\cvm
The Israel junction condition \re{2.6} is equivalent to
\begin{equation}\lae{2.11}
h_{\al\bet}=\ka(\hat T_{\al\bet}-\tfrac1n \hat T\bar g_{\al\bet}+\tfrac\s{n}\bar
g_{\al\bet}),
\end{equation}
where $\hat T=\hat T^\al_\al$.

\cvm
To solve these equations let us look at the embeddings of the form \re{0.9} or
\re{0.12}. Both forms are equivalent as we shall see later and both have their
advantages and disadvantages.

\cvm
Let us consider the embedding \re{0.9}, where $y=y(\tau,x^i)$ is parametrized in
the conformal time gauge. The induced metric is then of the form \re{0.10} and we
shall assume that the components of $(\hat T_{\al\bet})$ only depend on $\tau$.
Since $(\hat T_{\al\bet})$ is derived from perfect fluids, we deduce
\begin{equation}\lae{2.12}
\hat T^0_0=-\hat\rho,\q \hat T^\al_i=\hat p\msp \de^\al_i,
\end{equation}
where
\begin{equation}
\hat\rho=\rho +\sum_{k=0}^{k_0}\rho_k,\q \hat p= p+\sum_{k=0}^{k_0}p_k,
\end{equation}
and
\begin{equation}
p_k=\frac{\om_k}n\rho_k,\q \om_k\in\Z,
\end{equation}
\begin{equation}
\rho=-\tfrac12\norm{D\f}^2+V,
\end{equation}
and
\begin{equation}
p=-\tfrac12\norm{D\f}^2-V.
\end{equation}

Note that in the conformal time gauge
\begin{equation}
-\norm{D\f}^2=g\si{AB}\dot\f^A\dot\f^B e^{-2f}\equiv \norm{\dot\f}^2 e^{-2f}.
\end{equation}
We do not distinguish different norms by notation, since it will be evident by their
arguments how the norms are to be understood.

\cvm
Due to our assumption that the $(\sideset{_{k}}{}{\operatorname{{\it
T_{\al\bet}}}})$ are divergence free, we conclude
\begin{equation}
\rho_k=\bar\rho_ke^{-(n+\om_k)f},
\end{equation}
where $0\le \bar \rho_k$ is an integration constant, \cf, e.g., \cite[Lemma
0.2]{cg:brane}. Hence we have
\begin{equation}
\begin{aligned}\lae{2.19}
\hat\rho=\rho+\sum_{k=0}^{k_0}\bar\rho_k
e^{-(n+\om_k)f}=\rho+\sum_{k=0}^{k_0}\bar\rho_k (-r)^{-(n+\om_k)}.
\end{aligned}
\end{equation}

We observe that this relation remains valid, if we parametrize the embedding with
respect to $r$, the radial gauge, since $r=r(\tau)$ and $\tfrac{dr}{d\tau}\ne 0$, as
we shall show.

\br\lar{2.1}
We stipulate that $\om_0=n$ and that formally it should always be present, its
actual presence depending on the values of $\bar\rho_0$, namely, $\bar\rho_0>0$
or $\bar\rho_0=0$. Moreover, we assume $\om_k<n$ for $1\le k$.
\er

So far, we haven't used the Israel junction condition. Combining \re{2.11}, \re{2.12}
and \re{2.19} we infer
\begin{equation}\lae{2.20}
\begin{aligned}
h_{ij}=\tfrac\ka{n}(\hat\rho+\s)\bar g_{ij}&=\tfrac\ka{n}
(\rho+\s+\sum_{k=0}^{k_0}\bar \rho_k(-r)^{-(n+\om_k)})\bar g_{ij}\\[\cma]
&\equiv\tfrac\ka{n} (\rho+\s+\mc C)\bar g_{ij}
\end{aligned}
\end{equation}
for the spatial components of the second fundamental form.

\cvm
Let us now derive $h_{ij}$ from the Gau{\ss} formula in the radial gauge. In this
gauge, $N$ is a graph in the variables $(r,x^i)$ and we suppose that $\Om$ lies
below the brane. In this case, the normal vector
$\nu$ in the Gau{\ss} formula, which is supposed to point  to the exterior of $\Om$,
must have a positive component with respect to the $t$ coordinate.

The induced metric $(\bar g_{\al\bet})$ is given by \re{0.13}, the tangent vector is
equal to 
$y'=(1,t',0,\ldots,0)$, and the exterior normal
\begin{equation}
\nu=(\nu^\al)=\frac{\tilde h^{-\tfrac12}}{\sqrt{1-\tilde h^2\abs{t'}^2}}(\tilde h^2
t',1,0,\ldots,0).
\end{equation}

From \re{2.4} we conclude
\begin{equation}
\begin{aligned}
y^t_{ij}&=-\cha ij0 t' +\tilde\C^t_{ab}y^b_iy^c_j\\[\cma]
&=-h_{ij}\nu^t=-h_{ij} \frac{\tilde h^{-\tfrac12}}{\sqrt{1-\tilde h^2\abs{t'}^2}},
\end{aligned}
\end{equation}
where $(\cha \al\bet\ga)$ \resp $(\tilde\C ^c_{ab})$ are the Christoffel symbols in
$N$ \resp $\mc N$, and we deduce further
\begin{equation}
h_{ij}=\frac{\tilde ht'}{\sqrt{\tilde h^{-1}-\tilde h\abs{t'}^2}}\msp r^{-1}\bar g_{ij},
\end{equation}
and hence
\begin{equation}\lae{2.24}
\frac{\tilde ht'}{\sqrt{\tilde h^{-1}-\tilde h\abs{t'}^2}}=\tfrac\ka{n} (\rho+\s+\mc
C)r,
\end{equation}
in view of \re{2.20}.

Setting
\begin{equation}
a=\abs{\tilde h}^2\abs{t'}^2
\end{equation}
yields
\begin{equation}\lae{2.26}
\begin{aligned}
\frac1{1-a}&=\frac a{1-a}+1=\tfrac{\ka^2}{n^2}(\rho+\s+\mc C)^2r^2 \tilde
h^{-1}+1\\[\cma]
&=\tilde h^{-1}(\tilde h+\tfrac{\ka^2}{n^2}(\rho+\s+\mc C)^2r^2)
\end{aligned}
\end{equation}
or
\begin{equation}\lae{2.27}
\tilde h^{-1}(1-a)=\frac1{\tilde h+\tfrac{\ka^2}{n^2}(\rho+\s+\mc C)^2r^2}.
\end{equation}
We need this relation for future reference.

\cvm
Now, in the radial gauge, we have
\begin{equation}
\begin{aligned}
\rho &=-\tfrac12\norm{D\f}^2+V=\tfrac12\norm{\dot\f}^2\frac1{\tilde
h^{-1}-\tilde h\abs{t'}^2}+V\\[\cma]
&=\tfrac12\norm{\dot\f}^2\frac{\tilde h}{1-a}+V
\end{aligned}
\end{equation}
from which, together with relation \re{2.24}, we infer
\begin{equation}
a(1-a)=\tfrac{\ka^2}{4n^2} (\norm{\dot\f}^2+2 \tilde h^{-1}(1-a)(V+\s+\mc
C))^2 r^2 \tilde h.
\end{equation}

The solutions to this quadratic equation are
\begin{equation}\lae{2.30}
\begin{aligned}
a& =\frac{\tilde h+2(V+\s+\mc
C)^2\tfrac{\ka^2}{n^2}r^2+\norm{\dot\f}^2\tfrac{\ka^2}{n^2}\tilde h
r^2(V+\s+\mc C)}{2(\tilde h+\tfrac{\ka^2}{n^2}(V+\s+\mc C)^2r^2)}\\[\cma]
&\qq\pm \frac{\tilde h\sqrt{1-2\norm{\dot\f}^2\tfrac{\ka^2}{n^2}(V+\s+\mc
C)r^2 -\tfrac{\ka^2}{n^2}\norm{\dot\f}^4\tilde hr^2}}{2(\tilde
h+\tfrac{\ka^2}{n^2}(V+\s+\mc C)^2r^2)}.
\end{aligned}
\end{equation}
The solution with the plus sign will be the right one in our case, as we shall show later.

With that particular solution we further conclude
\begin{equation}\lae{2.31}
\begin{aligned}
-\tfrac12\norm{D\f}^2&=\tfrac12\norm{\dot\f}^2\frac{\tilde h}{1-a}\\[\cma]
&=\frac{\norm{\dot\f}^2\fried}{1-\tfrac{\ka^2}{n^2}\norm{\dot\f}^2(V+\s+\mc
C)r^2-D},
\end{aligned}
\end{equation}
where
\begin{equation}\lae{2.32}
D=\dis.
\end{equation}

\cvm
This quantity as well as $-\tfrac12\norm{D\f}^2 r^{2n}$ are of particular interest; we
shall find solutions such that both terms are positive in the limit
\begin{equation}\lae{2.33}
\lim_{r\ra 0} -\tfrac12\norm{D\f}^2r^{2n}>0,
\end{equation}
and
\begin{equation}
\lim_{r\ra 0}\dis>0.
\end{equation}

\cvm
Moreover, we shall only be interested in solutions satisfying
\begin{equation}\lae{2.35}
\lim_{r\ra0}Vr^{2n}=0.
\end{equation}

Define the vector field
\begin{equation}
\chi^A=\dot\f^Ar^{-(n-1)}
\end{equation}
and let
\begin{equation}
\begin{aligned}
\hat h&=\tilde h r^{4n-2}=(-r)^{3n-1}\tilde h(-r)^{n-1}\\[\cma]
&=(-r)^{3n-1}(m+\tfrac2{n(n+1)}\Lam(-r)^{n+1}-\tilde\ka(-r)^{n-1}).
\end{aligned}
\end{equation}

With these definition we infer from \re{2.31}
\begin{equation}
\begin{aligned}
-\tfrac12\norm{D\f}^2r^{2n}&=\frac{\norm{\chi}^2(\hat
h+\tfrac{\ka^2}{n^2}(V+\s+\mc C)^2r^{4n})}
{1-\tfrac{\ka^2}{n^2}\norm{\chi}^2(V+\s+\mc C)r^{2n}-\hat D},
\end{aligned}
\end{equation}
where
\begin{equation}\lae{2.39}
\hat D=\sqrt{1-2\tfrac{\ka^2}{n^2}\norm{\chi}^2(V+\s+\mc
C)r^{2n}-\tfrac{\ka^2}{n^2}\hat h\norm{\chi}^4}.
\end{equation}

Abbreviate the denominator as 
\begin{equation}
1-u-\sqrt{1-2u-v},
\end{equation}
then we conclude
\begin{equation}
\begin{aligned}
(1-u-&\sqrt{1-2u-v})(1-u+\sqrt{1-2u-v})\\[\cma]
&= (1-u)^2-(1-2u-v)=u^2+v,
\end{aligned}
\end{equation}
and hence
\begin{equation}
1-u-\sqrt{1-2u-v}=(u^2+v)(1-u+\sqrt{1-2u-v})^{-1}.
\end{equation}

\cvm
By definition
\begin{equation}
v=\kn \norm{\chi}^4\hat h,
\end{equation}
\begin{equation}
u^2=\tfrac{\ka^4}{n^4}\norm{\chi}^4(V+\s+\mc C)^2r^{4n}
\end{equation}
and thus
\begin{equation}
u^2+v=\tfrac{\ka^2}{n^2}\norm{\chi}^4(\hat h+\tfrac{\ka^2}{n^2}(V+\s+\mc
C)^2r^{4n})
\end{equation}
and we conclude
\begin{equation}\lae{2.45}
-\tfrac12\norm{D\f}^2 r^{2n}=\frac{1-\tfrac{\ka^2}{n^2}\norm{\chi}^2(V+\s+\mc
C)r^{2n}+\hat D}{\tfrac{\ka^2}{n^2}\norm{\chi}^2},
\end{equation}
\cf \re{2.39}.

By assumption $\mc C$ is at most of order $r^{-2n}$, i.e.,
\begin{equation}
\lim_{r\ra 0}\mc C r^{2n}=\bar\rho_0,
\end{equation}
\cf \rr{2.1}.

The term $\hat D$ in  \re{2.39} has always to be positive, which will be the case
near the singularity, if
\begin{equation}\lae{2.47}
2\kn \norm{\chi(0)}^2\bar\rho_0<1,
\end{equation} 
where $\norm{\chi(0)}^2$ can be looked at either as an initial value or as a limit
\begin{equation}
\lim_{r\ra 0}\norm{\chi}^2,
\end{equation}
which will always be assumed to be positive.

Let us summarize what we have proved so far for $-\tfrac12\norm{D\f}^2r^{2n}$, or
equivalently, for $\rho r^{2n}$ in a lemma

\bl\lal{2.2}
Let $N$ be solution of the embedding problem, satisfying the estimates \re{2.33},
\re{2.35} and
\re{2.47}, or more precisely, such that the corresponding term $\hat D$ in \re{2.39}
is positive, then the quantity
$\rho r^{2n}$ is a smooth function in the (independent) variables
$(r,\chi^A,\f^B)$ according to the equation \re{2.45}. The smoothness is even valid
in a full neighbourhood, $-\e<r<\e$, of the singularity $r=0$, provided
$\norm{\chi}^2>0$.
\el

\cvm
Let us now transform the wave equation \re{2.10}. We are still in the radial gauge
and conclude from \re{2.3}
\begin{equation}
\begin{aligned}
\D\f^A&=\bar g^{\al\bet}\f^A_{\al\bet}=\bar g^{00}\f^A_{00}+\bar
g^{ij}\f^A_{ij}\\[\cma]
&=\bar g^{00}\f^A_{,00}-\bar\C^0_{00}\f^A_0\bar g^{00}-\bar
g^{ij}\cha ij0\f^A_0+\hat\C^A_{BC}\f^B_0\f^C_0\bar g^{00}\\[\cma]
&=\Ddot\f^A\bar g^{00}-\cha 000\dot\f^A\bar g^{00}-\bar g^{ij}\cha ij0\dot\f^A,
\end{aligned}
\end{equation}
where
\begin{equation}
\Ddot\f^A=\tfrac D{dr}\dot\f^A
\end{equation}
is the \tit{covariant} second derivative of $(\f^A)$ with respect to $r$.

Now,
\begin{equation}
-\bar g^{ij}\cha ij0 =\tfrac nr\bar g^{00},
\end{equation}
\begin{equation}
-\cha 000=\tfrac12 (\tilde h^{-1} -\tilde h\abs{t'}^2)'\bar g^{00}
\end{equation}
and
\begin{equation}
\bar g^{00}=-\frac1{\tilde h^{-1}-\tilde h\abs{t'}^2}=-\frac{1}{{\tilde
h^{-1}}(1-a)},
\end{equation}
in our previous notation.

Thus we have
\begin{equation}
\begin{aligned}
\D\f^A=(\Ddot\f^A+\tfrac nr\dot\f^A-\tfrac12\frac{(\tilde h^{-1}(1-a))'}{\tilde
h^{-1}(1-a)}\dot\f^A)\bar g^{00}=\frac{\pa V}{\pa \f\si A},
\end{aligned}
\end{equation}
or equivalently,
\begin{equation}\lae{2.55}
\begin{aligned}
\Ddot\f^A&=-\tfrac nr \dot\f^A+\tfrac12\frac{(\tilde h^{-1}(1-a))'}{\tilde
h^{-1}(1-a)}\dot\f^A-\frac{\pa V}{\pa \f\si A}(\tilde h^{-1}(1-a)).
\end{aligned}
\end{equation}

In view of \re{2.27} we have
\begin{equation}
\tilde h^{-1}(1-a)=\frac{r^{4n-2}}{\hat h+\kn (\rho r^{2n}+\s r^{2n}+\mc C
r^{2n})^2}
\end{equation}
and therefore
\begin{equation}
\begin{aligned}
&(\tilde h^{-1}(1-a))'=\frac{(4n-2)r^{4n-2}r^{-1}}{\friedc}\\[\cma]
& -\frac{r^{4n-2}(\hat h' +2\kn (\rho r^{2n}+\s r^{2n}+\mc C r^{2n})(\rho
r^{2n}+\s r^{2n}+\kn C r^{2n})')}{(\friedc )^2},
\end{aligned}
\end{equation}
i.e.,
\begin{equation}
\begin{aligned}
&\frac{(\tilde h^{-1}(1-a))'}{\tilde h^{-1}(1-a)}=(4n-2)r^{-1} \\[\cma]
& -\frac{\hat h'
+2\kn (\rho r^{2n}+\s r^{2n}+\mc C r^{2n})(\rho r^{2n}+\s r^{2n}+\kn C
r^{2n})'}{\friedc}.
\end{aligned}
\end{equation}

Finally, setting as before
\begin{equation}
\chi^A=\dot\f^Ar^{-(n-1)},
\end{equation}
such that
\begin{equation}
\dot\chi^A=\Ddot\f^Ar^{-(n-1)}-(n-1)\dot\f^Ar^{-n},
\end{equation}
we deduce from \re{2.55}
\begin{equation}\lae{2.60}
\begin{aligned}
\dot\chi^A&=-\frac{\hat h'
+2\kn (\rho r^{2n}+\s r^{2n}+\mc C r^{2n})(\rho r^{2n}+\s r^{2n}+\kn C
r^{2n})'}{\friedc}\chi^A\\[\cma]
&\hp{=}-\frac{\pa V}{\pa \f\si A} \,\frac{r^{3n-1}}{\friedc}.
\end{aligned}
\end{equation}

Furthermore, as we shall prove in \rl{3.4},
\begin{equation}
(\rho r^{2n})'=\frac d{dr}(\rho r^{2n})=2n V r^{2n-1},
\end{equation}
and hence, equation \re{2.60} can be written as
\begin{equation}\lae{2.62}
\begin{aligned}
\dot\chi^A&=-\frac{\hat h'
+2\kn (\rho r^{2n}+\mc {\hat C})(2nV r^{2n-1}+\mc{\hat
C}')}{\frcc}\chi^A\\[\cma] 
&\hp{=}-\frac{\pa V}{\pa \f\si A}
\,\frac{r^{3n-1}}{\frcc},
\end{aligned}
\end{equation}
where the term $\rho r^{2n}$ has to be replaced with the help of \re{2.45},
and where we used the abbreviation
\begin{equation}\lae{2.63}
\mc {\hat C}=(\s+\mc C)r^{2n}.
\end{equation}

\cvm
The equation \re{2.62} has the form
\begin{equation}\lae{2.64}
\tfrac D{dr}\chi ^A=F^A(r,\f,\chi),
\end{equation}
where $F=(F^A)$ is a smooth vector field which is defined in $(-\e,\e)\times
\mc S\times T^{1,0}(\mc S)$ as long as $\norm{\chi}^2> 0$,  the right-hand
side of \re{2.45} is well defined and positive, and $\e$ is small. 

A solution 
$\chi(r)$ is considered to be a vector field over the map $\f(r)$, i.e., if $\f(r)$ is not
known a priori, then the system is incomplete and has to be completed before a
solution can be found, \cf equation \re{2.70}.

\bl\lal{2.3}
The field equation \re{2.64} has to be satisfied by any solution of the embedding
problem, that satisfies the equations \re{2.10},  \re{2.24} and the specified
assumptions on radial symmetry, with $\chi=\dot\f r^{-(n-1)}$, provided
\begin{equation}\lae{2.65}
\norm{\chi(0)}^2=\lim_{r\ra 0}\norm{\chi}^2>0,
\end{equation}
\begin{equation}\lae{2.66}
\lim_{r\ra 0}\rho r^{2n}>\bar\rho_0,
\end{equation}
\begin{equation}\lae{2.67}
\lim_{r\ra 0}V r^{2n}=0,
\end{equation}
and  the estimate \re{2.47} is valid.  The last assumption is redundant, if
$\bar\rho_0=0$.
\el

\bp
In the derivation of the field equation \re{2.64} we had at one point to make a
choice, namely, to pick one of two possible solutions in \re{2.30}. If the above
assumptions would imply that the choice with the plus sign is the only possible, then
the lemma would be proved.

We shall compare the limit in \re{2.66} depending on the possible choices in
\re{2.30}. If we choose the plus sign, then the term $-\tfrac12\norm{D\f}^2r^{2n}$
has been explicitly expressed in \re{2.45} and, if we denote its limit by $\rho_+$, we
get
\begin{equation}\lae{2.68}
\rho_+=\frac{1-\kn\norm{\chi(0)}^2\bar\rho_0+\sqrt{1-2\kn
\norm{\chi(0)}^2\bar\rho_0}}{\kn \norm{\chi(0)}^2}.
\end{equation}

In view of the condition \re{2.47}, the estimate $\rho_+>\bar\rho_0$ is valid, as
one easily checks.

On the other hand, if the corresponding quantity $\rho_-$ is derived, we obtain
\begin{equation}
\rho_-=\frac{\kn \norm{\chi(0)}^2\bar\rho_0}{1-\kn\norm{\chi(0)}^2
\bar\rho_0+\sqrt{1-2\kn
\norm{\chi(0)}^2\bar\rho_0}}\,\bar\rho_0,
\end{equation}
where the quotient is less than or equal to 1, in view of \re{2.47}.
\ep

Let us conclude this section with the following theorem

\bt\lat{2.4}
The first order system
\begin{equation}\lae{2.70}
\begin{aligned}
\dot\chi^A&=F^A(r,\f,\chi)-\hat\C^A_{BC}\chi^B\chi^Cr^{n-1}\\[\cma]
\dot \f^B&=\chi^Br^{n-1}
\end{aligned}
\end{equation}
where the dot indicates an ordinary derivative with respect to $r$, has a smooth
solution $(\chi^A,\f^B)$ in a maximal interval $(-\e_1,\e_2)$, $\e_i>0$, for any
initial values $\chi(0)$, $\f(0)$, provided $\chi(0)$ is such that
$\norm{\chi(0)}^2>0$ and the estimate \re{2.47} is satisfied. Using the relation
\re{2.68} as the definition of $\rho_+$, we also define $\rho_+$ for the solution of
\re{2.70}.
\et

The first equation in \re{2.70} is just a rephrasing of \re{2.64}. The existence of a
smooth solution of the initial value problem  follows from the general existence
theorem for integral curves. Notice also that the assumption $\norm{\chi(0)}^2>0$
can be satisfied, since the metric of $\mc S$ has at least one positive eigenvalue.

\cvm
We shall show in the next section that, if we restrict the solution from \rt{2.4} to
$\{r<0\}$, then
 $\f$ is a solution of our embedding problem.

\section{Existence of a global solution of the embedding problem}\las{3}

By a \cq{global solution} of the embedding problem we mean a solution given by
\rt{2.4} that is a solution of the embedding problems in $\mc N$ \resp $\mc {\hat
N}$, when restricted to $(-\e,0)$ \resp $(0,\e)$, $\e>0$ small.

\cvm
We shall prove that the solution of \rt{2.4} has indeed the required properties, if $n$
and the $\om_k$ are odd, since it satisfies
\begin{equation}\lae{3.1}
\rho_+>\rho_0.
\end{equation}

Let us first show that the solution in \rt{2.4} is also a solution of the embedding
problem in $\mc N$.

\cvm
Look at the embedding  in the conformal time gauge $x^0=\tau$, \cf
\re{0.9} and \re{0.10}, where now, however, the value $\tau=0$ should be mapped
onto $r=-\de$, $\de>0$, i.e., we stay away from the singularity.

\cvm
We shall show that this problem can be solved in a maximal interval $J=(a,b)$, with
$0\in J$, for given initial values at $\tau=0$. If we then choose as initial values those
of the solution from
\rt{2.4} evaluated at $r=-\de$, where of course $\de<\e$, 
then we shall conclude that the solution given by \rt{2.4} and the actual solution of
the embedding problem coincide. 

\cvm
In view of the maximality of $J$, $b$ must then correspond  to $r=0$ and the
solution of \rt{2.4}, restricted to $(-\e,0)$, will indeed be a solution of the
embedding problem in $\mc N$.

\cvm
In the conformal time gauge the energy density $\rho$ can be expressed as
\begin{equation}
\begin{aligned}
\rho&=-\tfrac12\norm{D\f}^2+V \\[\cma]
&=\tfrac12\norm{\dot\f}^2e^{-2f}+V,
\end{aligned}
\end{equation}
where we again use the same symbol $\norm{\cdot}$ to denote different norms. A
dot or a prime now indicates differentiation with respect to $\tau$.

The Israel junction conditions lead to the so-called modified Friedman equation
\begin{equation}\lae{3.3}
\abs{f'}^2=\tilde h+\kn (\rho+\s+\mc C)^2e^{2f},
\end{equation}
\cf e.g., the derivation in \cite[Section 1]{cg:brane}. Here, $\mc C$ has the same
meaning as before.

We have to find $f=f(\tau)$ and a map $\f$ from $N$ to $\mc S$,
$\f=\f(\tau,x^i)$, such that the relations \re{3.3} and the wave equation \re{2.10}
are satisfied. Since $\f$ is supposed to depend only on $\tau$, an easy exercise
shows that
\begin{equation}
\D\f^A=-(\Ddot\f^A+(n-1)f'\dot\f^A)e^{-2f},
\end{equation}
where a dot now indicates covariant differentiation with respect to $\tau$. Hence we
obtain as second requirement
\begin{equation}\lae{3.5}
\Ddot\f^A=-(n-1)f'\dot\f^A-\frac{\pa V}{\pa\f\si A}e^{2f}.
\end{equation}

Setting
\begin{equation}\lae{3.6}
\tilde\chi^A=\dot\f^Ae^{(n-1)f}
\end{equation}
we conclude
\begin{equation}\lae{3.7}
\dot{\tilde\chi}=-\frac{\pa V}{\pa \f\si A} e^{(n+1)f}.
\end{equation}

Now we can solve the pair of equations \re{3.3} and \re{3.5}. Notice also that $f'$
has to be negative.

\bt\lat{3.1}
The equations \re{3.3} and \re{3.5} have a smooth solution $(f,\f^A)$ in the interval
$(-\e,0]$ for small $\e>0$, with arbitrary initial values
\begin{equation}\lae{3.8}
\begin{aligned}
f(0)&=\log(\de),\\[\cma]
\f^A(0)&=\f_0^A,\\[\cma]
\dot\f^A(0)&=\xi^A,
\end{aligned}
\end{equation}
provided $\norm{\xi}^2>0$.
\et

\bp
Consider the first order system
\begin{equation}\lae{3.9}
\begin{aligned}
\dot{\tilde\chi}&=-\frac{\pa V}{\pa\f\si
A}e^{(n+1)f}-\hat\C^A_{BC}\tilde\chi^B\tilde\chi^C e^{-(n-1)f},\\[\cma]
\dot\f^B&=\tilde\chi^B e^{-(n-1)f},\\[\cma]
f'&=-\sqrt{\tilde h+\kn (\rho +\s+\mc C)^2e^{2f}},
\end{aligned}
\end{equation}
where all derivatives are ordinary derivatives with respect to $\tau$, with initial
values in \re{3.8}, except that now
\begin{equation}
\tilde\chi^A(0)=\xi^A\de^{n-1}.
\end{equation}

This system has a smooth solution on $(-\e,0]$ for small $\e>0$.

The requirement $\norm{\xi}^2>0$ can be satisfied, since $(g\si{AB})$ is supposed
to have at least one positive eigenvalue.

\cvm
Now, to avoid confusion denote the solution from \rt{3.1} by $\tilde\f^A$ and the
solution from \rt{2.4} by $\f^A$ and $\chi^A$.

We want to choose the initial values in \re{3.8} such that
\begin{equation}
\tilde\f^A(0)=\f^A(-\de)
\end{equation}
and
\begin{equation}\lae{3.12}
\frac{d}{dr}\tilde\f^A(-\de)=\chi^A(-\de)(-\de)^{n-1},
\end{equation}
where we observe that $\tilde \f$ can also be viewed as depending on $r$, since
$f'$ is negative.

We have
\begin{equation}
\begin{aligned}
\frac{d}{dr}\tilde\f^A&=\frac d{d\tau}\tilde\f\,\frac{d\tau}{dr}=-\frac
d{d\tau}\tilde\f \,\frac{e^{-f}}{f'}\\[\cma]
&= \frac d{d\tau}\tilde\f\,\frac{e^-f}{\sqrt{\tilde h+\kn(\rho+\s+\mc C)^2 e^{2f}}},
\end{aligned}
\end{equation}
in view of \re{3.9}. The requirement \re{3.12} is equivalent to
\begin{equation}
\chi^A(-\de)(-\de)^{n-1}=\frac{\xi^Ae^{-f}}{\sqrt{\tilde h+\kn(\rho+\s+\mc C)^2
e^{2f}}}.
\end{equation}

Obviously, we can choose 
\begin{equation}
\xi^A=\lam \chi^A(-\de),\qq\lam\in\R[],
\end{equation}
so that we only need to fine-tune $\lam$ such that
\begin{equation}\lae{3.16}
\lam^2=\de^{2n}(\tilde h+\kn(\rho+\s+\mc C)^2\de^2).
\end{equation}

The quantity $\rho$ is equal to
\begin{equation}\lae{3.17}
\begin{aligned}
\rho&=\tfrac12 \norm{\dot\f}^2e^{-2f}+V\\[\cma]
&=\tfrac12 \lam^2\norm{\chi(-\de)}^2\de^{-2}+V.
\end{aligned}
\end{equation}

Combining \re{3.16} and \re{3.17} we conclude
\begin{equation}\lae{3.18}
\begin{aligned}
&\lam^2=2\de^{-2(n-1)}\norm{\chi(-\de)}^{-4}\tfrac{n^2}{\ka^2}
\Big(1-(V+\s+\mc C)\de^{2n}\kn\norm{\chi(-\de)}^2\\[\cma]
&+\sqrt{1-2(V+\s+\mc C)\de^{2n}\kn \norm{\chi(-\de)}^2
-\kn
\de^{4n-2}\norm{\chi(-\de)}^4\tilde h}\,\Big).
\end{aligned}
\end{equation}

Of course, the power law integration constants are the same for both solutions,
especially $\bar\rho_0$. Looking at the equation \re{2.68}, and observing that
\re{2.47} is valid,  we conclude, that there exists $\de_0>0$, such that the relation
\re{3.18} can be satisfied for any
$0<\de<\de_0$ with $\lam=\lam(\de)\in\R[]$.

\cvm
Let us now determine the value of $\rho\de^{2n}$ for our solution. Combining
\re{3.17} and \re{3.18} we immediately deduce that
\begin{equation}
\lim_{\de\ra 0}\rho\de^{2n}=\rho_+,
\end{equation}
where $\rho_+$ is the quantity that belongs to the solution from \rt{2.4}, i.e.,
$\rho_+>\bar\rho_0$. Hence \rl{2.3} is applicable and the pair
$(\bar\chi^A,\tilde\f^B)$, where
\begin{equation}
\bar\chi^A=\frac d{dr}\tilde\f^Ar^{-(n-1)},
\end{equation}
satisfies the field equation \re{2.64}, or equivalently, the system \re{2.70}.

Since $(\bar\chi^A,\tilde\f^B)$ coincide with $(\chi^A,\f^B)$ at $r=-\de$, we
deduce
\begin{equation}
(\chi^A,\f^B)=(\bar\chi^A,\tilde\f^B)
\end{equation}
in a neighbourhood of $r=-\de$. 

\cvm
Now, define
\begin{equation}
M=\set{r\in (-\e,0)}{\tilde\f^A(r)=\f^A(r)}.
\end{equation}

Since $M$ is not empty, and due to the assumption, that $\tilde\f^A$ should be
defined in a maximal interval, we easily deduce that $M=(-\e,0)$, i.e., the solution
given by \rt{2.4} is indeed a physical solution in $\mc N$.

\cvm
Now, assume that $n$ and the $\om_k$ are odd---we haven't used these
assumptions yet---, then, looking at the definition of the vector field $F^A(r,\f,\chi)$
in \re{2.62}, we see that, had we considered the first variation of the action integral
in $\mc {\hat N}$ with all the subsequent derivations, we would have obtained the
identical vector field $F^A$.

\cvm
Hence, the preceding arguments, valid in $\mc N$, also hold in $\mc {\hat N}$, and
we conclude further that the solution of \rt{2.4}, restricted to $(0,\e)$, is also a
physical solution in
$\mc {\hat N}$.

\cvm
Since the transformation
\begin{equation}\lae{3.23.1}
(r,t,x^i)\ra (r,-t,x^i)
\end{equation}
is an isometry in $\mc N$ as well as $\mc{\hat N}$, the previous considerations are
valid for upper branes as well as lower branes, since the isometry in \re{3.23.1} maps
an upper brane to a lower brane and vice versa.

Branes that are given by an embedding of the form
\begin{equation}
y(r,x^i)=(r,t(r),x^i)
\end{equation}
can be easily distinguished, as being an upper or lower brane, by looking at the
equation satisfied by $t'$.

An upper brane has to satisfy the equation \re{2.24}, while, in case of a lower brane,
the equation looks like
\begin{equation}\lae{3.25a}
\frac{\tilde ht'}{\sqrt{\tilde h^{-1}-\tilde h\abs{t'}^2}}=-\tfrac\ka{n} (\rho+\s+\mc
C)r,
\end{equation}
These equations are identical in $\mc N$ \resp $\mc{\hat N}$.

\cvm
Thus, let us express $t'$ in our case, where $N\su\mc N$ is an upper brane, by
combining \re{2.24} and \re{2.26} to deduce
\begin{equation}
t'=\tfrac\ka{n}\tilde h^{-1}\frac{(\rho +\s+\mc C)r}{\sqrt{\tilde h+\kn
(\rho+\s+\mc C)^2r^2}},
\end{equation}
or equivalently,
\begin{equation}\lae{3.23}
t'=-\tfrac\ka{n}\frac1{\tilde hr^{n-1}}\,\frac{(\rho r^{2n}+\mc{\hat C})r^{n-1}}
{\sqrt{\frcc}},
\end{equation}
\cf the definition in \re{2.63}.

Since $t=t(r)$ is a smooth function in $(-\e,\e)$ and the right-hand side of \re{3.23}
has constant sign in $\{r\ne 0\}$,  this implies that the brane $\hat
N\su\mc{\hat N}$ that corresponds to the big bang part of the global solution must
be a lower brane.

Let us summarize this observation in

\br\lar{3.2a}
Let $(N,\hat N)$ be a global solution of the embedding problem, then the branes $N,
\hat N$ have opposite positions, i.e., if $N$ is an upper brane, then $\hat N$ is a
lower brane, and vice versa.\qedhere
\er
\ep

\br\lar{3.2}
Let us emphasize that even in the general case, i.e., without assuming that $n$ and
the $\om_k$ are odd, 
$t=t(r)$ is of class
$C^\un((-\e_,0])$, since this is the case for $\f^A$ and $\chi^A$, as well as $\rho
r^{2n}$, because $\rho r^{2n}$ satisfies the differential equation
\begin{equation}\lae{3.24}
\frac d{dr}(\rho r^{2n})=2n V r^{2n-1},
\end{equation}
as we shall prove in
\er

\bl\lal{3.4}
Let $\f^A$ be a solution of the embedding problem, then $\rho r^{2n}$ satisfies
\re{3.24}.
\el

\bp
We choose the conformal time gauge. $(T_{\al\bet})$ is a divergence free perfect
fluid stress-energy tensor, hence $\rho$ satisfies
\begin{equation}\lae{3.29a}
\dot\rho=-n(\rho+p)f'=-2n(\rho-V)f',
\end{equation}
since
\begin{equation}
\rho+p=-\norm{D\f}^2=2(\rho-V),
\end{equation}
where dot or prime indicate differentiation with respect to $\tau$.

Hence
\begin{equation}
\frac{d\rho}{dr}=\dot\rho\frac1{rf'}=-2n(\rho-V)r^{-1}
\end{equation}
and
\begin{equation}
\frac d{dr}(\rho r^{2n})=2nVr^{2n-1}.\qedhere
\end{equation}
\ep

\br
The brane, the existence of which we have just proved, is of class $(B)$ with
$\tilde\ga=2n-1$, i.e.,
\begin{equation}\lae{3.29}
\lim_{\tau\ra b}\abs{f'}^2e^{2\tilde\ga f}=\tilde m>0,
\end{equation}
where $b\in\eR$ is such that
\begin{equation}
\lim_{\tau\ra b}e^f=0.
\end{equation}
Hence, $b$ is finite and we may choose $b=0$.
\er

\bp
See \cite[Definition 0.1]{cg:brane} for the definition of class $(B)$ spacetimes. The
conclusion \cq{$b$ finite} is proved in \cite[Lemma 3.1]{cg:arw}, so that it remains
to verify \re{3.29}.

\cvm
From \re{3.3} we conclude
\begin{equation}
\abs{f'}^2 e^{2\tilde\ga f}=\frcc,
\end{equation}
i.e.,
\begin{equation}
\lim_{\tau\ra b}\abs{f'}^2e^{2\tilde\ga f}=\kn(\lim_{r\ra 0}\rho
r^{2n}+\bar\rho_0)^2.\qedhere
\end{equation}
\ep

\section{The reflection method}

In case $\mc S$ is locally symmetric and $V$ invariant in the neighbourhood of a
point $p\in\mc S$ with respect to the geodesic symmetry centered at $p$, then the
big bang part $\hat N$ of the global solution can be obtained by an odd reflection of 
$N$, namely, define
\begin{equation}
\hat y(r,\hat t(r),x^i)=(r,-t(-r),x^i)
\end{equation}
for $r>0$, where we have normalized the globally defined function $t(r)$,
representing the embedding of $N$, if $r<0$, by setting $t(0)=0$, which can be
easily achieved by adding a constant.

\cvm
We want to prove that
\begin{equation}
\hat t(r)=t(r),\qq r>0,
\end{equation}
and hence conclude that $t$ is an odd function.

\bt\lat{4.1}
Let $n$ and $\om_k$, $0\le k\le k_0$, be odd and let $\mc S$ be a locally
symmetric, semi-Riemannian space such that its metric $(g\si{AB})$ has at least one
positive eigenvalue. Furthermore, assume that there is a point $p\in\mc S$ such that
$V\in C^\un (\mc S)$ is invariant in a neighbourhood of $p$ with respect to the
geodesic symmetry centered at $p$.

If we  choose in the global existence theorem, \rt{2.4}, the initial condition $\f(0)=p$,
then the quantity $\rho r^{2n}$  can be looked at as an even function  in a
neighbourhood of
$r=0$.

Moreover, introducing Riemannian normal coordinates around $p$, such that
$\f^A(0)=0$, the globally defined scalar fields map $(\f^A)$ is an odd map in a
neighbourhood of $r=0$. 

The function $t=t(r)$ is odd as well in a neighbourhood of
$r=0$.
\et

\bp
After the introduction of Riemannian normal coordinates around $p$, such that $p$
corresponds to $p=0$, the geodesic symmetry centered in $p$ is nothing but odd
reflection
\begin{equation}
\f^A\ra -\f^A,
\end{equation}
i.e., the invariance property of $V$ can be expressed  by saying that $V$ is even in
a neighbourhood of $\f=0$.

\cvm
The vector field $F^A(r,\f,\chi)$ is defined in \re{2.62}. Let us define a variant of
$F^A$
\begin{equation}
\F^A(r,u,\f,\chi)
\end{equation}
by treating the term $\rho r^{2n}$ in $F^A$ as a new variable $u$.

\cvm
We found our solution of the global embedding problem as a solution of the first order
system \re{2.70}, \cf \rt{2.4}.

Since the quantity $\rho r^{2n}$, evaluated for that particular solution, is smooth
and satisfies
\begin{equation}\lae{4.9}
\frac d{dr}(\rho r^{2n})=2n V r^{2n-1},
\end{equation}
the solution from \rt{2.4} will also be a solution of the modified system
\begin{equation}\lae{4.6}
\begin{aligned}
\dot\chi^A&=\F^A(r,u,\f,\chi)-\hat\C^A_{AB}\chi^B\chi^Cr^{n-1},\\[\cma]
\dot\f&=\chi^Br^{n-1},\\[\cma]
\dot u&=2n V r^{2n-1}
\end{aligned}
\end{equation}
to the given initial values.

Notice that we choose
\begin{equation}
u(0)=\lim_{r\ra 0}\rho r^{2n}.
\end{equation}

\cvm
 Let us denote the global solution of \re{4.6} by $\f^A(r)$,
$\chi^B(r)$, and $u(r)$.

\cvm
Now, define for $r>0$ the reflected quantities
\begin{equation}
\begin{aligned}
\tilde\chi^A(r)&=\chi^A(-r),\\[\cma]
\tilde\f^B(r)&=-\f^B(-r),\\[\cma]
\tilde u(r)&=u(-r),
\end{aligned}
\end{equation}
then
\begin{equation}
\begin{aligned}
\dot{\tilde\chi}^A(r)&=-\dot\chi^A(-r),\\[\cma]
\dot{\tilde\f}^B(r)&=\dot\f^B(-r),\\[\cma]
\dot{\tilde u}(r)&=-\dot u(-r).
\end{aligned}
\end{equation}

$\tilde u,\tilde\f,\tilde\chi$ are smooth in $[0,\e)$ and in $r=0$ they have the same
values as $u$, $\f$, $\chi$. If we can prove that $(\tilde u,\tilde\f,\tilde\chi)$
satisfy the first order system in \re{4.6} in $(0,\e)$, the uniqueness of the solution
will imply that they coincide with $(u,\f,\chi)$, and the theorem will be proved, in view
of \re{3.23}.

\cvm
Now, for a locally symmetric space, the Christoffel symbols $\hat\C^A_{BC}$ are
odd, i.e.,
\begin{equation}
\hat\C^A_{BC}(-\f)=-\hat\C^A_{BC}(\f)
\end{equation}
in Riemannian normal coordinate, and one easily checks that
\begin{equation}
\F^A(r, \tilde u,\tilde\f,\tilde\chi)=-\F^A(-r,u(-r),\f^A(-r),\chi^A(-r)).
\end{equation}

Hence $(\tilde u,\tilde\f,\tilde\chi)$ are solutions and the theorem is proved.
\ep

\section{Colliding branes}

Consider a domain $\Om\su\mc N$ bounded by two branes $N_1$ and $N_2$. If
$N_1$ is the upper brane and $N_2$ the lower, and if, after having solved the global
embedding problem for each single brane, resulting in branes $\hat N_1$ and $\hat
N_2$, we would want to describe the configuration in $\mc{\hat N}$ similarly as in
$\mc N$, namely, that $\hat N_i$ are boundary branes of a domain $\hat \Om$,
then this is only possible, if $N_1$ and $N_2$ collide at the singularity.

\bl
Suppose that each single brane $N_i$ can be considered as the big crunch part of a
globally defined solution, then the branes $\hat N_i$ are boundary branes of domain
$\hat \Om\su\mc {\hat N}$, if and only if the branes $N_i$ collide at the singularity. 
\el

\bp
Let us write the global solutions as globally defined graphs
\begin{equation}
t=u_i(r,x^i),\q \q r\in(-\e,\e),
\end{equation}
with smooth functions $u_i$. 

Without loss of generality let us assume that $N_1$ is an upper brane and $N_2$ a
lower, i.e., we have
\begin{equation}
u_1\ge u_2,\qq r<0.
\end{equation}
However, in $\mc{\hat N}$ the branes reverse positions and therefore we have
\begin{equation}
u_1\le u_2,\qq r>0,
\end{equation}
and hence $u_1(0)=u_2(0)$.

\cvm
On the other hand, assume that $u_1(0)=u_2(0)$. Since $\Om$ can be written as
\begin{equation}
\Om=\set{(r,t,x^i)}{u_1(r)> t> u_2(r), r<0},
\end{equation}
we immediately see that 
\begin{equation}
\hat\Om=\set{(r,t,x^i)}{u_1(r)< t< u_2(r), r>0}.\qedhere
\end{equation}
\ep

\br
If the assumptions of part (iii) of \rt{0.1} are satisfied for two boundary branes, then
it can be easily achieved that they collide at the singularity by translating one brane
in the $t$-direction.
\er

\cvb
\noindent

%\fbox{%

\begin{minipage}{5cm}
\begin{center}
\begin{mfpic}[20]{-2}{2}{-2}{2}
%\headshape{1}{1}{false}
\axis{y}
\store{aa}{\function{-1.5,1.5,0.1}{(-0.7*x**3)}}
\store{ab}{\function{-1.5,1.5,0.1}{(0.7*x**3)}}
\pointdef{A}(0,0)
\pointfilltrue
\point{\A}
\drawcolor{blue}\mfobj{aa}
\tlabel(-1,-0.5){$\Om$}
\tlabel(1,0.5){$\hat\Om$}
%\drawcolor{red}\mfobj{ab}
%\tcaption{\raggedright{\it Figure 1:}  The reflection of a single brane}
\end{mfpic}%

\cvm
{\larger[-1] A globally defined single brane}.
\end{center}
\end{minipage}
%}
\hbox{}\hfill
\begin{minipage}{5cm}
\begin{center}
\begin{mfpic}[20]{-2}{2}{-2}{2}
%\headshape{1}{1}{false}
\axis{y}
\store{ac}{\function{-1.8,1.8,0.1}{(0*x**3)}}
%\drawcolor{blue}\mfobj{aa}
\drawcolor{red}\mfobj{ac}
%\drawcolor{red}\mfobj{ab}
\pointdef{A}(0,0)
\pointfilltrue
\point{\A}
\tlabel(-1.5,-0.5){$\Om$}
\tlabel(1.5,0.5){$\hat\Om$}
%\tcaption{\raggedright{\it Figure 1:}  The reflection of a single brane}
\end{mfpic}%

\cvm
{\larger[-1] The  totally geodesic brane $\{t=\const\}$.}
\end{center}
\end{minipage}
\hfill

\cvb
Below are two examples of  colliding branes. The $t$-axis represents the black (white) hole singularity.

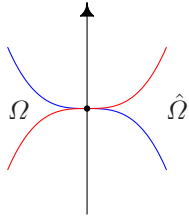
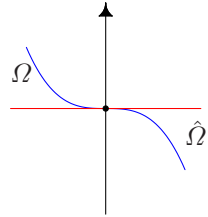
\cvb
\begin{minipage}{5cm}
\begin{center}
\begin{mfpic}[20]{-2}{2}{-2}{2}
%\headshape{1}{1}{false}
\axis{y}
%\store{aa}{\function{-1.5,1.5,0.1}{(-0.7*x**3)}}
%\store{ab}{\function{-1.5,1.5,0.1}{(0.7*x**3)}}
\drawcolor{blue}\mfobj{aa}
\drawcolor{red}\mfobj{ab}
\pointdef{A}(0,0)
\pointfilltrue
\point{\A}
\tlabel(-1.5,-0.2){$\Om$}
\tlabel(1.5,-0.2){$\hat\Om$}
%\tcaption{\raggedright{\it Figure 1:}  The reflection of a single brane}
\end{mfpic}%
\end{center}
\end{minipage}
%\hskip 2cm
\hfill
\begin{minipage}{5cm}
\begin{center}
\begin{mfpic}[20]{-2}{2}{-2}{2}
%\headshape{1}{1}{false}
\axis{y}
%\store{aa}{\function{-1.5,1.5,0.1}{(-0.7*x**3)}}
\store{ac}{\function{-1.8,1.8,0.1}{(0*x**3)}}
\drawcolor{blue}\mfobj{aa}
\drawcolor{red}\mfobj{ac}
\pointdef{A}(0,0)
\pointfilltrue
\point{\A}
\tlabel(-1.8,0.5){$\Om$}
\tlabel(1.5,-0.7){$\hat\Om$}
%\tcaption{\raggedright{\it Figure 1:}  The reflection of a single brane}
\end{mfpic}%
\end{center}
\end{minipage}
\hfill

\cvb

\cvm
From a physical and geometrical point of view it would be desirable to find sufficient
conditions guaranteeing that the branes collide only at the singularity.

\bt
Let $N_i$ be two branes in the black hole region of $\mc N$, which are boundary
branes of a domain $\Om$, where we assume that $N_1$ is the upper brane and
$N_2$ the lower brane. Parametrizing both branes with respect to $r<0$, we can
write each brane as a graph
\begin{equation}
t=u_i(r)
\end{equation}
such that
\begin{equation}
u_1\ge u_2.
\end{equation}
In view of the Israel junction conditions, the $u_i$ satisfy the equations
\begin{equation}
\frac{\tilde hu_i'}{\sqrt{\tilde h^{-1}-\tilde h\abs{u_i'}^2}}=\pm\tfrac\ka{n}
(\rho_i+\s_i+\mc C_i)r,
\end{equation}
\cf \re{3.25a}, where the plus sign corresponds to $u_1$ and the minus sign to
$u_2$.

Suppose, moreover, that the equation
\begin{equation}\lae{5.7}
(\rho_1+\s_1+\mc C_1)+(\rho_2+\s_2+\mc C_2)=0
\end{equation}
cannot be valid in the interval $(-a,0)$, then the only possible collision point in
$(-a,0]$ of the branes is $r=0$.
\et

\bp
This is an immediate consequence of Rolle's
theorem.
\ep

\section{Maximal extension of the branes}

We now consider branes without assuming that $n$ or the $\om_k$'s are odd, and
look at a solution given by \rt{2.4}, which, when
restricted to
$\{r<0\}$, is defined on a maximal interval $\{r_1<r\le 0\}$, with $-r_1<r_0$,
where $r_0$ is the black hole radius. 

Notice that the field equations \re{2.70} are
probably only valid in a much smaller interval
$\{-\e<r\le 0\}$, since the quantity $\hat D$ in \re{2.39} may no longer be positive
 outside the smaller interval. 

\cvm
We want to prove that we may choose $r_1=-r_0$, if $\mc S$ is Riemannian and
complete, and if in addition $\rho$ is bounded from below
\begin{equation}\lae{6.1}
\rho\ge -c,\qq c>0,
\end{equation}
which is slightly better than requiring that the potential $V$ is bounded from below. 

\cvm
Pick $\bar r\in (r_1,0)$, $\abs{\bar r}<1$, and switch to the conformal time gauge.
Set
\begin{equation}
\de =\log (-\bar r)<0,
\end{equation}
so that the brane is defined on a maximal interval $J=\{a<\tau\le\de\}$, $a\in \eR$.

The relevant terms are now $(f,\f^A)$ satisfying the equations \re{3.3} and
\re{3.5}, or the equivalent one \re{3.7}, which is 
\begin{equation}\lae{6.3}
\dot{\tilde\chi}=-\frac{\pa V}{\pa \f\si A} e^{(n+1)f},
\end{equation}
where
\begin{equation}\lae{6.4}
\tilde\chi^A=\dot\f^Ae^{(n-1)f}.
\end{equation}
All derivatives are covariant derivatives with respect to $\tau$, \cq{$\tfrac
D{d\tau}$}.

We certainly will hit a coordinate singularity, if we approach the horizon, i.e., $e^f$
will always be bounded from below and above by positive constants. The quantity
$\mc C$ in \re{3.3} is likewise uniformly bounded.

\cvm
For convenience let us also change from $\tau$ to $-\tau$, so that $J=[\de, a)$,
$0<\de$; notice that this switch does not effect \re{6.3} and \re{6.4}, because
these equations are tensor equations. The only difference is that $f'>0$.

\bl
Let $J=[\de, a)$ and assume that horizon is not reached, then $a<\un$.
\el

\bp
Since $f'>0$, we deduce from \re{3.3}
\begin{equation}\lae{6.5}
f'\ge \sqrt{\tilde h}\ge c>0,
\end{equation}
since the horizon is not reached. Hence, $a<\un$, for otherwise $f$ would tend to
infinity and so would $r=-e^f$.
\ep

\bl
If $a<\un$ and \re{6.1} is valid, then
\begin{equation}
\norm{\dot\f}\le c\qq\A\,\tau\in J,
\end{equation}
$\f(J)\su\mc S$ is precompact, and hence
\begin{equation}
\abs{V}\le c\q \text{and}\q \abs{\rho}\le c.
\end{equation}
\el

\bp
(i) We use the relation \re{6.3} to conclude
\begin{equation}
\begin{aligned}
V(\de)-V(\tau)&=-\int_\de^\tau V'=-\int_\de^\tau \frac{\pa V}{\pa \f\si
A}\dot\f^A\\[\cma]
&= \int_\de^\tau(\tfrac12\norm{\tilde\chi}^2)'e^{-2nf}\\[\cma]
&=\tfrac12\norm{\tilde\chi}^2e^{-2nf}
\Big|^\tau_\de+n\int_\de^\tau\norm{\tilde\chi}^2f'.
\end{aligned}
\end{equation}

Now, in view of \re{6.5} and the definition of $\rho$, we infer
\begin{equation}
c\int_\de^\tau\norm{\dot\f}^2+\rho(\tau)\le \rho(\de),
\end{equation}
and hence
\begin{equation}
\int_\de^\tau\norm{\dot\f}^2\le c\qq\A\,\tau \in J.
\end{equation}

Moreover, because of the equation \re{3.29a} there holds
\begin{equation}
\dot\rho=-n(\tfrac12\norm{\dot\f}^2e^{-2f})f'\le 0,
\end{equation}
i.e., $\rho$ is bounded from above and thus we have
\begin{equation}\lae{6.12}
\abs{\rho}\le c\qq\A\,\tau\in J.
\end{equation}

\cvm
(ii) Next we want to show that $\f:J\ra \mc S$ is uniformly H\"older continuous. Let
$d$ be the Riemannian distance in $\mc S$, then, for any $\de\le\tau\le\tau'<a$
\begin{equation}
\begin{aligned}
d(\f(\tau),\f(\tau'))&\le \int_\tau^{\tau'}\norm{\dot\f}\le
\abs{\tau'-\tau}^{\tfrac12} \Big(\int_\tau^{\tau'}\norm{\dot\f}^2\Big)^{\tfrac12}
\\[\cma]
&\le c\abs{\tau'-\tau}^{\tfrac12},
\end{aligned}
\end{equation}
i.e., 
\begin{equation}
\f\in C^{0,\tfrac12}(\bar J,\mc S),
\end{equation}
 and $\f$ can be extended to $\bar J$ as a H\"older continuous function, since $\mc
S$ is complete. But $\bar J$ is compact, and hence $\f(\bar J)$ is compact, and we
conclude that $V\circ\f$ is uniformly bounded on $J$, and hence $\norm{\dot\f}$, in
view of \re{6.12}.
\ep

But with these a priori estimates it follows immediately that the equations \re{3.3}
and \re{3.5} can be solved on a larger interval, i.e., $J$ wouldn't be maximal, a
contradiction, which can only be resolved by assuming that 
\begin{equation}
\lim_{\tau\ra a}e^f=r_0.
\end{equation}

\nocite{steinhardt:cyclic,seiberg:transition}

\bibliographystyle{hamsplain}
\bibliography{mrabbrev,publications}

%\listoffigures

%\cleardoublepage
\closegraphsfile
%\thispagestyle{empty}
%\closegraphsfile
\end{document}